\documentclass[journal]{IEEEtran}
\usepackage{graphicx}
\newcommand{\sarrow}{\mathrel{\scalebox{0.75}{$\rightarrow$}}}
\usepackage{array}
\usepackage{makecell}
\newcolumntype{C}[1]{>{\centering\arraybackslash}m{#1}}
\usepackage{booktabs}
\usepackage{multirow}
\usepackage{amsmath}
\usepackage{amssymb}
\usepackage{bm}
\usepackage{makecell}
\usepackage[binary-units=true]{siunitx}
\usepackage{siunitx}
\usepackage{enumitem}
\usepackage{xcolor}

\usepackage{cite}
\usepackage[caption=false,font=footnotesize]{subfig}

\usepackage[
    colorlinks=true,
    linkcolor=black,
    citecolor=black,
    urlcolor=black
]{hyperref}

\newcolumntype{C}[1]{>{\centering\arraybackslash}p{#1}}

\title{On-Device Super-Resolution Imaging for a Low-Cost SPAD Array on a 640-KB-SRAM Microcontroller}

\author{Zhenya~Zang,~\IEEEmembership{}
        Istvan~Gyongy$^{\dagger}$,~\IEEEmembership{Senior Member,~IEEE,}
        ~and Mike~Davies$^{\dagger}$,~\IEEEmembership{Fellow,~IEEE}
\thanks{$^{\dagger}$These authors jointly supervised this work and 
        are co-senior authors.
        The authors are with the School of Engineering,
        University of Edinburgh, Edinburgh EH9 3FF, U.K.
        (e-mail: zzang@ed.ac.uk; igyongy2@exseed.ed.ac.uk; Mike.Davies@ed.ac.uk).} 
}

\begin{document}

\maketitle
\begin{abstract}
This work presents a super-resolution (SR) deep-learning (DL) architecture to simultaneously upscale low-resolution (LR) depth and intensity images acquired by a consumer-grade VL53L9CX time-of-flight (ToF) sensor, a compact single-photon avalanche diode (SPAD) ranging device, while targeting deployment on a resource-constrained microcontroller (MCU) with only \SI{640} KB SRAM and 2~MB Flash. The VL53L9CX provides  $54\times42$ depth and intensity measurements. The proposed framework performs $\times4$ spatial SR to reconstruct $216\times168$ depth and intensity images simultaneously. The network employs separate reconstruction branches for intensity and high-resolution (HR) depth. We train the network with efficient loss functions and investigate the performance by testing multiple compact, hardware-oriented backbones, including Spatially-Adaptive Feature Modulation (SAFM), Swift Parameter-Free Attention Network (SPAN), and Residual Local Feature Network (RLFN). We evaluate the network on both synthetic and real measurements, with particular attention to constrained activation and memory storage. We export the selected network to ONNX, quantize it to INT8, generate C inference code for an STM32H563ZI Arm Cortex-M33 MCU, and integrate it with our simplified sensor's firmware. This work is the first to demonstrate a compact DL SR model on a low-cost bare-metal MCU for a consumer-grade LR SPAD sensor, integrated with customized MCU firmware to enable on-device SR inference.
\end{abstract}


\section{Introduction}
\IEEEPARstart{D}{irect} time-of-flight (dToF) sensing is increasingly demanded in robotic systems, autonomous platforms, and three-dimensional (3-D) perception because it directly measures photon ToF with nano- and picosecond temporal resolution. In single-photon avalanche diode (SPAD)-based dToF systems, SPADs provide high single-photon sensitivity, while precise on-chip time-to-digital converters (TDCs) \cite{henderson2019192} or histogramming circuitry \cite{gyongy2023direct} enable accurate depth estimation from photon-arrival statistics. Compared with indirect time-of-flight (iToF) approaches that measure the phase shift of modulated illumination, dToF enables longer-range operation, higher temporal resolution, and more robust detection of multiple returns. Wide-field solid-state light detection and ranging (LiDAR) systems combine SPAD arrays with integrated timing circuits, enabling parallel data acquisition without mechanical scanning, thereby reducing system complexity and sensitivity to mechanical vibration. Custom-made, research-grade SPAD systems typically need to be synchronized with a laser source costing anywhere from hundreds to tens of thousands of US dollars (USD). In addition to research-grade SPAD imagers and high-end industrial LiDAR systems, compact consumer-grade multi-zone dToF sensors have recently become increasingly accessible. Their spatial resolution has also improved substantially, progressing from early $8\times8$ ranging-zone arrays \cite{martin2021all} to $20\times15$ \cite{kappel2024bsi_tof}, $32\times48$ devices \cite{ams_tmf8829_2025}, and recently up to $54\times42$ ranging zones \cite{st_vl53l9cx_2026}. All sensors integrate a laser source and data transfer interface, with a sensor-module price under 65 USD. Despite increasing spatial resolution and low-power, on-device histogram processing and image generation, it remains difficult to achieve resolutions comparable to conventional RGB or infrared cameras because limited on-chip memory constrains fine temporal resolution, long-range measurements, and generating higher-resolution (HR) transient histograms from many pixels, thereby limiting computer-vision applications.

 Data-driven super-resolution (SR) deep learning (DL) models provide feasible methods to recover HR spatial information from input low-resolution (LR) images. However, most modern SR architectures are developed, trained, and executed on GPUs, smartphones, or embedded processors with more memory than conventional microcontrollers (MCUs), which have only hundreds of KB of SRAM and several MB of Flash. In this work, we investigate depth and intensity SR imaging for a VL53L9CX SPAD array under the severe computational resource and memory constraints of an STM32H563ZI MCU. The target platform contains a single-core Arm Cortex-M33 CPU operating at 250 MHz, 2 MB Flash memory, and only \SI{640}  KB SRAM. The primary challenge is not only reducing the number of neural network parameters (stored in the Flash) but also managing intermediate activation outputs, which dominate SRAM consumption, especially as spatial dimensions increase during up-scaling modules. Consequently, network design must also consider firmware integration, as memory should be allocated for other data, such as sensor calibration and LR and HR image storage.

\begin{table*}[!t]
\centering
\caption{Comparison of super-resolution and depth-enhancement methods
using the compact SPAD sensors.}
\label{tab:related_work}
\scriptsize
\setlength{\tabcolsep}{3.0pt}
\renewcommand{\arraystretch}{0.8}
\begin{tabular}{@{}
C{2.65cm}
C{1.35cm}
C{2.05cm}
C{1.85cm}
C{0.85cm}
C{2.75cm}
C{2.15cm}
C{1.95cm}
@{}}
\toprule

\makecell[c]{\textbf{Work}} &
\makecell[c]{\textbf{ToF}\\\textbf{Sensor}} &
\makecell[c]{\textbf{LR $\sarrow$ HR}} &
\makecell[c]{\textbf{Training}\\\textbf{datasets}} &
\makecell[c]{\textbf{On-}\\\textbf{MCU}} &
\makecell[c]{\textbf{Inference}\\\textbf{Platform}} &
\makecell[c]{\textbf{Platform}\\\textbf{Memory}} &
\makecell[c]{\textbf{Precision /}\\\textbf{Speed (fps)}} \\

\midrule
\multicolumn{8}{c}{\textbf{Fusion-based methods}} \\
\midrule

Pixels2Pose$^a$ (2022) \cite{2022pixels2pose} &
VL53L5CX &
\makecell{$4{\times}4$ $\sarrow$ $32{\times}32$} &
Histogram + RGB &
No &
NVIDIA Tesla RTX 6000 &
24 GB VRAM &
FP32 / 31.3 \\

DELTAR (2022) \cite{li2022deltar} &
VL53L5CX &
\makecell{$8{\times}8$ $\sarrow$ $480{\times}640$} &
Histogram + RGB &
No &
-- &
-- &
FP32 / -- \\

CFPNet (2025) \cite{2025cfpnet} &
VL53L5CX &
\makecell{$8{\times}8$ $\sarrow$ $480{\times}640$} &
Histogram + RGB &
No &
\makecell{NVIDIA RTX 2080 Ti} &
11 GB VRAM/GPU &
FP32 / -- \\

DEPTHOR (2025) \cite{2025depthor} &
VL53L5CX &
\makecell{$8{\times}8$ $\sarrow$ $480{\times}640$} &
Histogram + RGB &
No &
\makecell{NVIDIA RTX 3090} &
24 GB VRAM/GPU &
FP32 / -- \\

Nadalini \textit{et al.} (2026) \cite{nadalini2025multimodal} &
VL53L5CX &
\makecell{$8{\times}8$ $\sarrow$ $48{\times}48$} &
Depth + RGB &
Yes &
\makecell{GreenWaves\\GAP9 MCU} &
\makecell{1.5 MB SRAM\\64 MB Flash} &
BF16 / -- \\

LiteSense (2026) \cite{2026litesense} &
VL53L8CX &
\makecell{$8{\times}8$ $\sarrow$ $480{\times}480$} &
Histogram + RGB &
No &
\makecell{NVIDIA RTX 3060 \\ Intel i5-12400F \\ Rockchip RK3576 NPU} &
\makecell{8/12 GB VRAM \\ -- \\ 16 GB LPDDR4} &
\makecell{FP32 / 66.7 (GPU) \\ FP32 / 6.25 (CPU) \\ FP32 / 1.75 (NPU)} \\

\midrule
\multicolumn{8}{c}{\textbf{Non-fusion methods}} \\
\midrule

Wilson \textit{et al.} (2025) \cite{wilson2025oe} &
VL53L8CX &
\makecell{$4{\times}4$ $\sarrow$ $32{\times}32$} &
ToF parameters &
No$^{b}$ &
\makecell{NVIDIA RTX 3090} &
24 GB VRAM &
INT8$^{b}$ / -- \\

Zang \textit{et al.} (2025) \cite{zang2025oe} &
VL53L8CX &
\makecell{$8{\times}8$ $\sarrow$ $50{\times}50$} &
Depth + intensity &
No &
\makecell{NVIDIA RTX 1000 \\ Ada Generation} &
6 GB GDDR6 &
INT8 / 68.8 \\

Li \textit{et al.} (2025) \cite{2025facial} &
VL53L8CX &
\makecell{$8{\times}8$ $\sarrow$ $32{\times}32$} &
Histogram &
No &
-- &
-- &
-- / -- \\

Zang \textit{et al.} (2026) \cite{zang2025oe} &
TMF8829 &
\makecell{$32{\times}48$ $\sarrow$ \\ $256{\times}256$} &
Depth + intensity &
Yes$^{c}$ &
\makecell{Qualcomm QRB2210 SoC \\ Debian Linux} &
2 GB RAM &
INT8 / 5.8 \\

\textbf{This work (2026)} &
\textbf{VL53L9CX} &
\makecell{$\mathbf{54{\times}42}$ $\sarrow$ \\
          $\mathbf{216{\times}168}$} &
\textbf{Depth + intensity} &
\textbf{Yes} &
\makecell{\textbf{STM32H563ZI} \\
          \textbf{Cortex-M33 / bare metal}} &
\makecell{\textbf{\SI{640}  KB SRAM} \\
          \textbf{2 MB Flash}} &
\makecell{INT8 / 1.5 (I/D), \\ 0.7 (I+D)$^{d}$} \\

\bottomrule
\end{tabular}

\vspace{1mm}
\raggedright
\scriptsize
$^{a}$Pixel2Pose uses histograms and RGB input during training to estimate pose; SR occurs in the training objective, and the network outputs a skeleton pose rather than SR images. \\
$^{b}$Wilson \textit{et al.} discuss a projected INT8 MCU implementation; the MCU deployment and corresponding performance are not experimentally demonstrated. \\
$^{c}$The Arduino UNO Q contains an STM32U585 MCU and a Qualcomm QRB2210 SoC; neural SR inference is executed on the QRB2210 SoC rather than the MCU. \\
$^{d}$I/D denotes a single-branch output for either intensity or depth, while I+D denotes a two-branch output that reconstructs depth and intensity simultaneously.
\end{table*}

The main contributions of this work are as follows:
\begin{itemize}[leftmargin=*]
\item A compact DL model is trained using efficiently combined loss functions and high-fidelity dToF synthetic datasets covering various scenes for joint SR of depth and intensity measurements acquired using a VL53L9CX SPAD ToF sensor and processed on an STM32H563ZI evaluation board, with a total system cost of approximately USD 120.
\item Comprehensive offline evaluation of accuracy and model complexity for depth and intensity reconstruction is performed using both synthetic and real-world data, with extensive ablation studies, such as weight parameters and loss functions. Multiple DL backbones are trained and evaluated.
\item The DL model is developed using a bandwidth-friendly architecture with simplified PixelShuffle modules for upscaling. We then develop an automated quantization and MCU implementation pipeline for the DL model, along with simplified MCU firmware to ensure compact integrability with the embedded SR model and enable on-device SR imaging.
\end{itemize}

\section{Related Work}
Table \ref{tab:related_work} summarizes existing studies using compact SPAD arrays. The first stream is sensor fusion that leverages the high spatial resolution of multiple camera types. Representative works \cite{2022pixels2pose, li2022deltar, 2025cfpnet, 2025depthor} fuse histograms and HR RGB images as input during training, employ sophisticated model architectures, and are trained and evaluated on GPUs with tensor cores and sufficient memory. LiteSense \cite{2026litesense} runs its SR model on several computing platforms, including a high-performance embedded neural processing unit with 16 GB of memory. A recent work \cite{nadalini2025multimodal} implemented a compressed monocular depth-estimation model on a GAP9 MCU, where $8\times8$ depth measurements from a VL53L5CX sensor were used to generate pseudo-labels for on-device fine-tuning of a $48\times48$ RGB-to-depth model. An earlier monocular depth-estimation model \cite{peluso2021monocular} was deployed on an STM32F7 MCU, requiring about 337 KB of RAM and 651~ms per inference ($\sim$1.54~fps) for a $48\times48$ RGB input and producing a $48\times48$ depth map; however, it takes RGB rather than LR depth measurements as input.

Another category is the non-fusion strategies that take only LR images from a single sensor, without cross-camera image co-registration from multiple cameras. Another consumer-grade VL53L8CX SPAD array \cite{zang2025oe} uses an SR DL model, but it was demonstrated offline without MCU integration. An LR image-input-only DL model for both depth and intensity images has demonstrated imaging with a low-cost SPAD array TMF8829 at a higher spatial resolution of 48$\times$32 on embedded platforms \cite{zang2026device}. That implementation runs inference on an embedded system with about 2~GB of memory and operating-system support.  Two other works \cite{wilson2025oe}, \cite{2025facial} reported SR DL models that use pixel-wise histogram parameters and 3--D histogram information for SR imaging, but were not implemented on an embedded platform. Other work has used research-grade SPAD arrays that capture LR photon-counting (intensity) images alongside LR depth on a single sensor, and has applied deep learning \cite{martin2022_oe_video, ruget2021_HIST_Net} and optimization algorithms \cite{gyongy2020high} to guide depth upsampling. This review motivates our work, as fully on-MCU, non-fusion SR solutions for compact SPAD-based ToF sensors remain lacking. This work targets a bare-metal STM32H563ZI MCU with only 640 KB of SRAM mainly for activations and firmware data and 2 MB mainly for learned parameters. To the best of our knowledge, we present the first non-fusion SR DL model running entirely on a resource-constrained MCU and reconstructing $216\times168$ depth and intensity maps directly from $54\times42$ measurements.

\begin{figure*}[t]
  \centering
  \includegraphics[width=\textwidth]{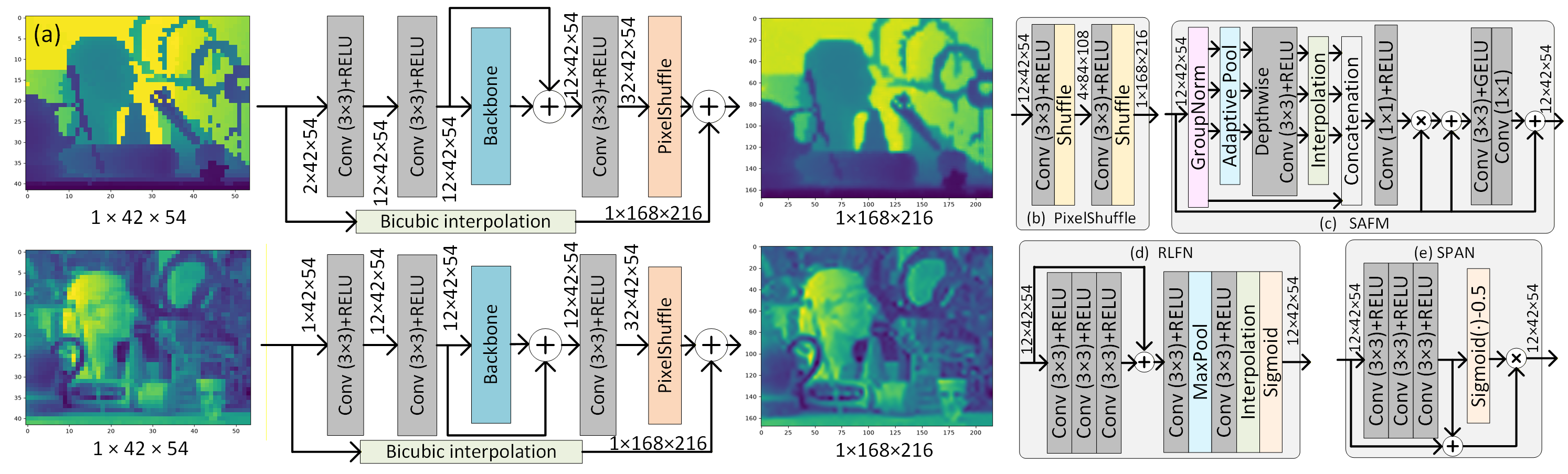}
    \caption{(a) DL architecture. Modality-specific branches process depth and intensity, with intensity-gradient guidance provided to the depth branch. (b) Simplified PixelShuffle module. The interchangeable backbone is (c) SAFM, (d) RLFN, or (e) SPAN. Each branch upsamples through two compact PixelShuffle stages and uses an interpolated residual connection for reconstruction.}
  \label{fig:DL_overview}
\end{figure*}

Recent DL model-to-MCUs research has demonstrated that convolutional neural networks can run on resource-constrained MCUs through integer quantization, static memory planning, and optimized kernels. MCUNet \cite{lin2020mcunet} and MCUNetV2 \cite{lin2021memory} introduce DL model and inference-runtime co-design for deploying memory-constrained MCUs, primarily for image classification and related vision tasks. Software toolchains such as TensorFlow Lite Micro \cite{david2021tensorflow}, CMSIS-NN \cite{lai2018cmsis}, and STM32Cube.AI \cite{stm32cubeai} provide quantized operators and hardware-specific kernels to support efficient MCU inference. However, SR is a challenging workload for resource-constrained deployment because dense HR intermediate features produce large activation tensors and consume substantial SRAM. Therefore, deploying an SR network on a sub-megabyte MCU requires not only simplified model design and quantization, but also careful control of feature-map dimensions and activation-memory allocation.

\section{Method}
\subsection{Network Architecture}
Suppose the LR depth and intensity measurements are
$D_{\mathrm{LR}}, I_{\mathrm{LR}} \in \mathbb{R}^{42\times54}$.
For an upsampling factor $s=4$, the reconstruction network
$f_{\theta}$ generates
\begin{equation}
    (\hat{D}_{\mathrm{HR}},\hat{I}_{\mathrm{HR}})
    = f_{\theta}(D_{\mathrm{LR}},I_{\mathrm{LR}}),
    \;
    \hat{D}_{\mathrm{HR}},\hat{I}_{\mathrm{HR}}
    \in \mathbb{R}^{168\times216}.
\end{equation}
where $\theta$ denotes the learnable network parameters. Fig.~\ref{fig:DL_overview} presents an overview of the DL architecture. The model uses two-modality reconstruction branches as depth and intensity encode different properties, where depth primarily represents scene geometry, whereas intensity reflects returned optical signal strength. Separate feature-processing paths limit the propagation of intensity texture into the reconstructed depth. The two branches are optimized jointly using
\begin{equation}
  \mathcal{L} = \mathcal{L}_{d} + \lambda_{i}\,\mathcal{L}_{i},
  \label{eq:total}
\end{equation}
where $\lambda_{i}$ balances the intensity and depth branches. For depth, we use the Charbonnier loss
\begin{equation}
  \mathcal{L}_{\mathrm{char}}(x, y)
    = \frac{1}{N}\sum_{p}\sqrt{\left(x_{p}-y_{p}\right)^{2}+\epsilon^{2}},
  \qquad \epsilon = 10^{-3},
  \label{eq:charb}
\end{equation}
a smooth and continuously differentiable function at all points~\cite{lai2017Charbonnier}. As boundary fidelity is crucial for depth images, the depth loss adds a first-order term that matches Sobel gradients between the prediction $\hat{D}$ and the ground truth $D$,
\begin{equation}
  \mathcal{L}_{\nabla}(\hat{D}, D)
    = \left\| \nabla_{x}\hat{D} - \nabla_{x}D \right\|_{1}
    + \left\| \nabla_{y}\hat{D} - \nabla_{y}D \right\|_{1},
  \label{eq:grad}
\end{equation}
which penalizes edge displacement rather than only the pointwise residual. A total-variation (TV) regularizer $\mathcal{L}_{\mathrm{TV}}$ suppresses high-frequency artifacts on planar surfaces, where the network otherwise injects texture that the metric depth map does not need, and a structural term
$\mathcal{L}_{\mathrm{SSIM}} = 1 - \mathrm{SSIM}(\hat{D}, D)$ penalizes loss of local structure. The depth objective becomes
\begin{equation}
  \begin{split}
    \mathcal{L}_{d} = {}& \mathcal{L}_{\mathrm{char}}(\hat{D}, D)
      + w_{\nabla d}\mathcal{L}_{\nabla}(\hat{D}, D) \\
      &+ w_{\mathrm{TV}}\mathcal{L}_{\mathrm{TV}}
      + w_{\mathrm{SSIM},d}\mathcal{L}_{\mathrm{SSIM}}(\hat{D}, D).
  \end{split}
  \label{eq:depth}
\end{equation}
The gradient and total variance (TV) terms act in opposition, the former sharpening transitions and the latter smoothing them, so $w_{\nabla d}$ and $w_{\mathrm{TV}}$ are tuned jointly. The intensity branch uses the same Charbonnier base with its own gradient and structural terms, but omitting TV as reflectance texture is signal rather than artifact for intensity,
\begin{equation}
  \mathcal{L}_{i}
    = \mathcal{L}_{\mathrm{char}}(\hat{I}, I)
    + w_{\nabla i}\mathcal{L}_{\nabla}(\hat{I}, I)
    + w_{\mathrm{SSIM},i}\mathcal{L}_{\mathrm{SSIM}}(\hat{I}, I).
  \label{eq:inten}
\end{equation}
For basic feature extraction modules, we evaluate three lightweight backbones under an identical main model structure. This isolates the backbone's influence while keeping the input and output interfaces identical. Lightweight backbones reduce parameters and multiply-accumulate (MAC) operations while maintaining reconstruction performance. The Spatially-Adaptive Feature Modulation (SAFM) Network ~\cite{SAFMN_2023_ICCV} (Fig. \ref{fig:DL_overview}(c)) splits the channels into four groups, pools each at a different scale, applies a depthwise convolution, and recombines them into a gate that multiplies the input. The adaptive pooling block provides different receptive fields while using fewer trainable parameters than convolutions. The Residual Local Feature Network (RLFN)~\cite{RLFN_2022_CVPR} (Fig.~\ref{fig:DL_overview}(d)) stacks three $3\times3$ convolutions with a residual connection, then rescales the result with a spatial attention mask formed by a pooling, a convolution, and a sigmoid. The Swift Parameter-free Attention Network (SPAN)~\cite{SPAN_2024_CVPR} shown in Fig. \ref{fig:DL_overview}(e) stacks three $3\times3$ convolutions and derives its attention map directly from their output through a fixed activation. The attention therefore costs no parameters. Nevertheless, a low parameter count alone does not guarantee compatibility with MCU-class hardware. Intermediate feature maps, unsupported operators, and quantization boundaries may dominate deployment cost.

The proposed network must satisfy hardware memory constraints, namely $M_{\mathrm{weights}} \leq M_{\mathrm{Flash}}$ and $M_{\mathrm{activation}} + M_{\mathrm{firmware}} \leq M_{\mathrm{SRAM}}$, where the target platform with single Arm Cortex-M33 CPU core, provides $M_{\mathrm{Flash}} = 2~\mathrm{MB}$ and $M_{\mathrm{SRAM}} = 640~\mathrm{KB}$. In our DL model, weight storage is not the dominant constraint. Instead, peak SRAM occupancy is primarily determined by intermediate feature maps, especially large activation tensors after the backbone. We modify the upscaling modules as they dominate memory usage. Let $r$ be the upscaling factor and $c$ the channel width entering the PixelShuffle; the convolution preceding it is $\mathrm{Conv}(c, c\,r^{2})$. The shuffle costs no arithmetic or weights, but it is memory-demanding because it rearranges rather than reduces. With $c=12$ and $r=4$ on a $42\times54$ input, the conventional single-stage head is $\mathrm{Conv}(12,192)$ followed by one $\times4$ shuffle. Its output, $192\times42\times54$, and the shuffled result, $12\times168\times216$, totally generate \SI{850} KB of features in INT8, more than the \SI{640} KB of SRAM. Splitting the one-stage $\times4$ into two-stage $\times2$ shuffles caps the first stage at $12\times2^{2}=48$ channels and the last at $1\times2^{2}=4$. We decrease the intermediate channels from 12 to 8, reducing the first stage to $8\times2^{2}=32$. The largest pair is then the first shuffle, $32\times42\times54$ in and $8\times84\times108$ out, achieving \SI{142} KB activation together ($5.99\times$ smaller than \SI{850} KB). The parameter count also reduces. The single-stage head is $\mathrm{Conv}(12,192)$ plus the $\mathrm{Conv}(12,1)$ that follows the shuffle, totaling 21037 INT-8 numbers. The two-stage head is $\mathrm{Conv}(12,32)$ plus $\mathrm{Conv}(8,4)$, totaling 3780 INT-8 numbers (a $5.6\times$ reduction).

\begin{figure}[t]
  \centering
  \includegraphics[width=\columnwidth]{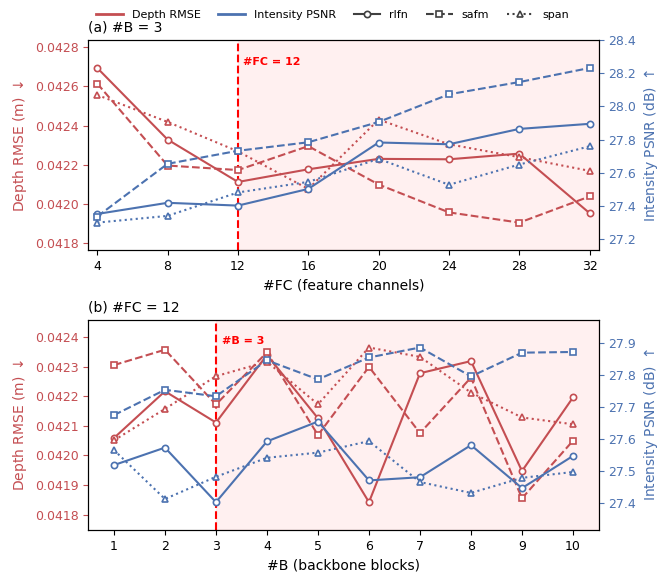}
  \caption{Depth RMSE (red, left) and intensity PSNR (blue, right) results of different \#FC and \#B. (a) feature channels \#FC at \#B$\,=\,3$ and (b) backbone blocks \#B at \#FC$\,=\,12$, on Middlebury test scenes. Line style marks the three backbones. The red dashed line marks the select configuration, where shading marks settings that exceeds the available SRAM, tested on the MCU.
  }
  \label{fig:dim_sweep_short}
\end{figure}

Model size is set by two dimension parameters: the number of feature channels (\#FC), shown as 12 in Fig.~\ref{fig:DL_overview}(a), and the number of blocks (\#B) inside the backbone, which that figure draws once rather than repeated. Figure~\ref{fig:dim_sweep_short} sweeps both against reconstruction quality on the Middlebury test scenes for all three backbones. Panel~(a) fixes \#B$=3$ and varies \#FC; panel~(b) fixes \#FC$=12$ and varies \#B. The red dashed line marks the deployed configuration, beyond which the model no longer fits the available memory. Intensity PSNR improves steadily with \#FC across the swept range, while \#B changes neither metric beyond the spread between individual runs, so \#FC is the parameter worth spending memory on. Since \#B has little effect on performance over the three models, we fix it at 1 for the remaining experiments.

\subsection{Training Strategy}
We implemented and trained all models in PyTorch on an NVIDIA RTX 5070 Ti 16 GB GPU using paired LR and HR depth and intensity images. We randomly split the dataset into 85\% for training and 15\% for validation subsets and applied horizontal and vertical flipping during training for data augmentation. We stacked LR depth and intensity maps into a two-channel $54\times42$ input and used the corresponding $216\times168$ depth and intensity maps as the ground truth (GT). All backbones were trained for 200 epochs with a batch size of 32 using the AdamW optimizer, with an initial learning rate of $1\times10^{-4}$, a weight decay of $1\times10^{-3}$, and momentum coefficients of $\beta=(0.9,0.99)$. We linearly warmed up the learning rate over the first five epochs, then reduced it using a cosine decay schedule with a minimum learning-rate ratio of $5\times10^{-4}$. We applied gradient clipping with a maximum norm of 1.0 to stabilize training. The depth reconstruction loss combines Charbonnier, SSIM, TV, and Sobel-gradient terms with weights $1$, $0.05$, $1\times10^{-4}$, and $0.2$, respectively. The intensity loss combines Charbonnier, SSIM, and Sobel-gradient terms with weights $1$, $0.1$, and $0.05$, respectively. The Charbonnier loss uses $\epsilon=10^{-3}$. Model checkpoints are saved when both validation SSIM and PSNR improve.

\subsection{Model Implementation and Firmware Customization}
\label{ssec:FW_N_SW}

\begin{table}[t]
  \centering
  \setlength{\tabcolsep}{3.0pt}
  \renewcommand{\arraystretch}{0.8}
  \scriptsize
  \caption{Deployment cost of the DL model using one of three INT8 quantized backbones on the MCU, at \#FC=12, \#B=1, and $\times4$ upscaling. MACs and FLOPs are counted over one forward pass. RAM and Flash figures are measured from the built firmware. We measure inference latency on the GPU and average over 20 iterations across the eight scenes.}
  \label{tab:models_size_split}
  \begin{tabular}{lcccccc}
    \toprule
    Backbone & \makecell{Params\\(K)} & \makecell{MACs\\(M)} & \makecell{FLOPs\\(M)}
             & \makecell{RAM usage\\(KB)} & \makecell{Flash usage\\(KB)} & \makecell{Latency\\(ms)} \\
    \midrule
    SAFM & 13.7 & 47.5 & 95.0  & 533.14 (83.3\%) & 170.96 (8.3\%)  & 0.75 ($\pm$0.03) \\
    SPAN & 14.3 & 51.3 & 102.5 & 523.05 (81.7\%) & 141.62 (6.9\%)  & 0.44 ($\pm$0.01) \\
    RLFN & 14.9 & 52.6 & 105.1 & 522.59 (81.7\%) & 148.86 (7.3\%)  & 0.62 ($\pm$0.01) \\
    \bottomrule
  \end{tabular}
\end{table}

ST's reference firmware allocates memory for depth, reflectance, ambient, and confidence outputs and contains multiple processing stages, including extraction, distance calibration, temporal noise reduction, confidence estimation, rate normalization, reflectance estimation, radial-to-perpendicular conversion, and subsequent depth-processing stages. Because the DL model takes only depth and intensity as inputs, we retain three processing kernels required to generate these inputs while reducing latency and SRAM consumption: extraction, distance calibration, and radial-to-perpendicular conversion. The latter two stages calibrate and correct the depth measurements, while the sensor output provides raw intensity measurements directly. The final depth-processing stage writes the processed depth directly into the network input tensor. This avoids approximately \SI{67.7} KB of additional SRAM allocation, including \SI{35.4} KB of calibration maps required by the omitted rate-normalization, reflectance, and maximum-detectable-distance processing, \SI{17.7} KB for the reflectance and confidence output frames, and \SI{14.5} KB for the second raw frame buffer used in ST's double-buffered acquisition. The retained preprocessing kernels require \SI{3.6}{\milli\second} per frame, compared with \SI{19.5}{\milli\second} for the original processing path, corresponding to a $5.4\times$ reduction in processing time. Including sensor acquisition and readout, acquisition with the reduced preprocessing pipeline takes approximately \SI{57}{\milli\second} per frame. Joint depth-and-intensity inference takes \SI{1425}{\milli\second}, whereas separately exported and compiled single-branch depth- or intensity-only models require approximately \SI{667}{\milli\second} per inference.

Table~\ref{tab:models_size_split} compares the deployment cost of the three quantized backbones. Although SAFM has the fewest parameters and computational overhead, with 13.7K parameters and 47.5M MACs, it exhibits the highest latency of \SI{1.72}{\milli\second}, compared with \SI{0.79}{\milli\second} for SPAN and \SI{1.40}{\milli\second} for RLFN. This indicates that the MAC count does not determine the practical execution time. In particular, SAFM has more complex operations beyond standard convolutions, including slicing, resizing, pooling, concatenation, element-wise operations, and tensor-layout transformations. These operations contribute barely to MAC but add memory access, tensor manipulation, and increased latency. Regarding memory footprint on an MCU, RLFN requires the least RAM, closely followed by SPAN, whereas SPAN requires the least Flash. With competitive reconstruction accuracy, SPAN shows a favorable efficiency--accuracy trade-off, while SAFM is retained for the subsequent quantization study due to the best reconstruction metrics.

\begin{figure}[t]
  \centering
  \includegraphics[width=\columnwidth]{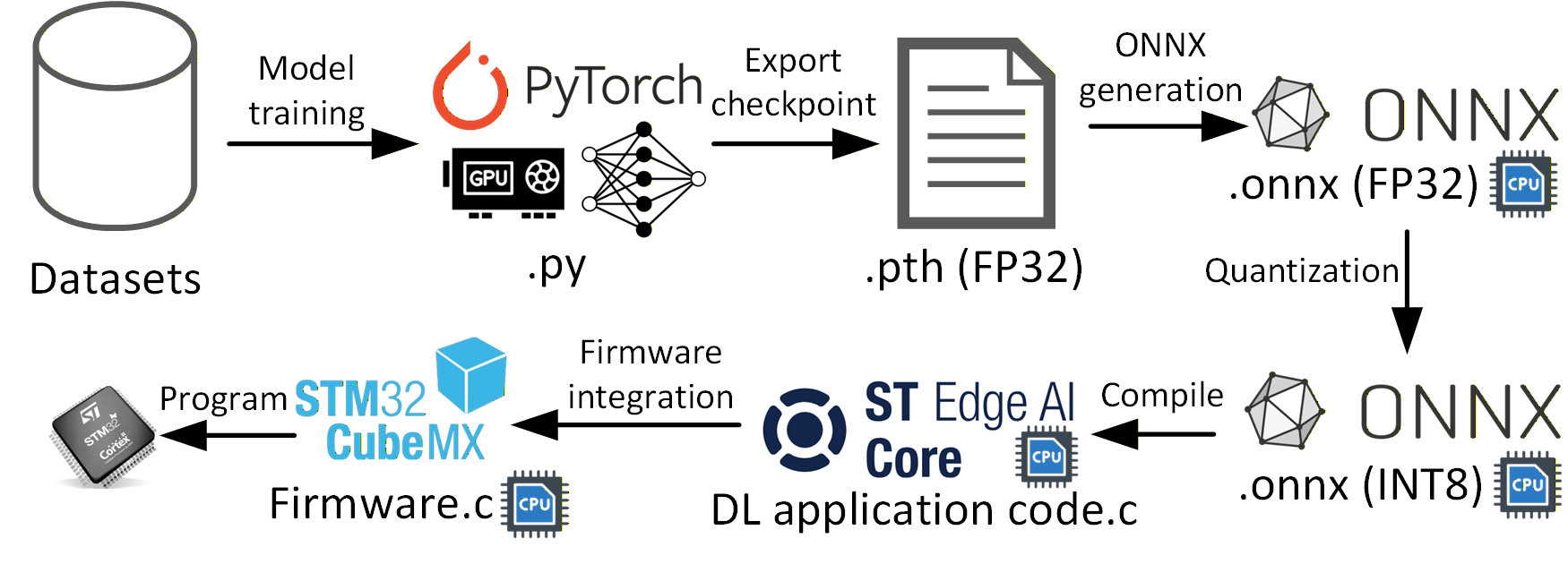}
  \caption{Deployment pipeline. The model is trained in PyTorch, exported to ONNX for static memory planning, and quantized to INT8. ST Edge AI Core compiles the quantized graph into C sources, which are integrated with the sensor firmware and programmed onto the STM32H563ZI.}
  \label{fig:deployment}
\end{figure}

\begin{figure}[t]
  \centering
  \includegraphics[width=\columnwidth]{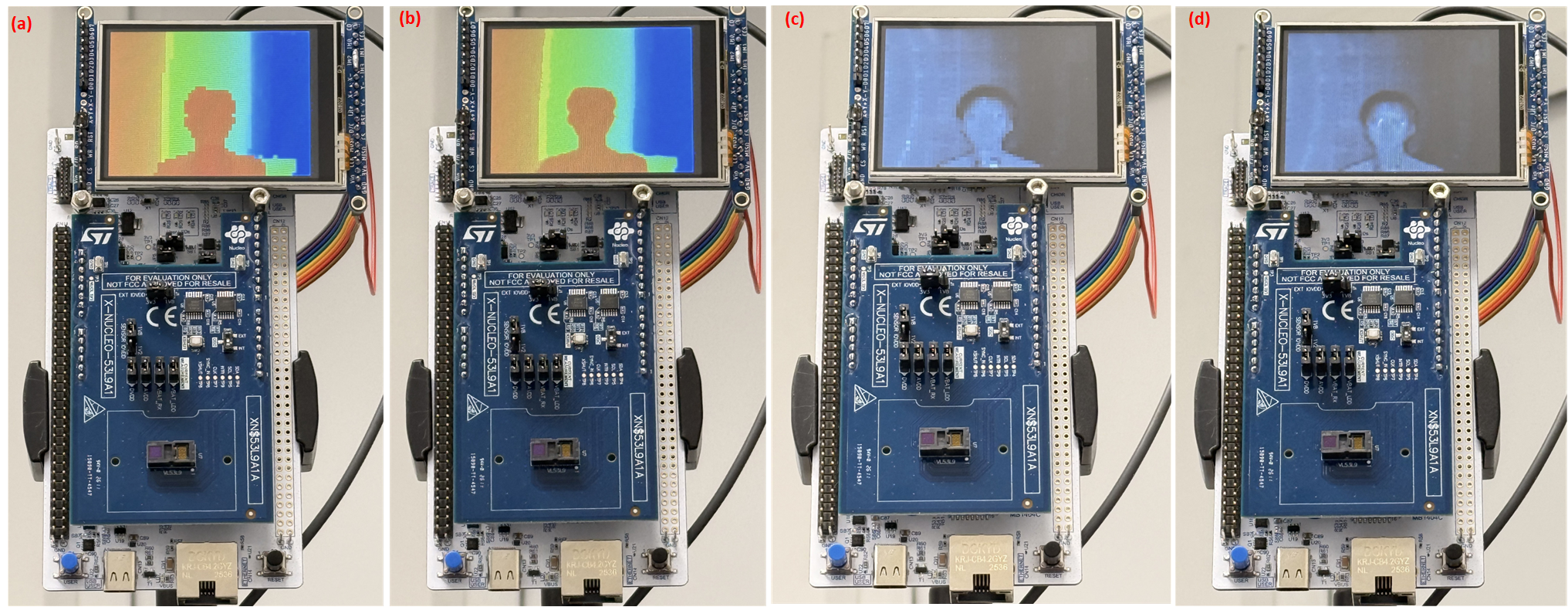}
  \caption{Assembled prototype platform comprising the MCU evaluation board, sensor board, and LCD. (a, b) Color-coded depth maps in LR and SR-HR display modes, respectively, showing a human silhouette against a background wall with continuously varying depth. (c, d) Corresponding grayscale intensity images in LR and SR-HR modes, respectively.}
  \label{fig:system}
\end{figure}

Fig. \ref{fig:deployment} shows the model-to-MCU implementation flow. After training on the GPU, we export the checkpoint to ONNX \cite{onnxruntime} with fixed input dimensions, and then quantize the FP32 model to INT8 in Quantize-De-Quantize (QDQ) form using post-training static quantization. ST Edge AI Core \cite{stedgeaicore} compiles the quantized model into C, where weights become an array in Flash, activations are continuously stored in SRAM whose size is fixed at compile time, and the network is declared with a generated header specifying the input and output tensor shapes, the quantization scales, and the required RAM size. The C application code receives the normalized sensor data in the input tensor and calls the generated inference function. The system-level frame rate depends on the readout method. Data for offline visualization on a PC is transferred over UART at 5 Mbaud, giving a per-frame latency of approximately 149 ms, while the image can also be displayed on an LCD over a 15.6 MHz SPI, achieving $\sim$79 ms per HR frame. Fig. \ref{fig:system} shows the imaging platform assembling the LCD and the MCU evaluation board, where (a) and (b) show LR and SR-HR depth videos; (c) and (d) show intensity counterparts. On the host, a receiver script reads the serial data and reassembles frames into the firmware's regularized format. Each transfer packet includes the LR depth and intensity maps, followed by the SR output. We reduced the heap reservation from \SI{128} KB to \SI{92} KB, leaving more SRAM for the network activation buffer.

\section{Experiments}
\subsection{Dataset}
Training pairs of LR and HR intensity and depth are generated from NYU~Depth~v2 \cite{nyudepthv2}, whose depth is captured with a structured-light camera, where object edges are blurred and a filtering step \cite{jeon2014intrinsic} is needed to fill missing depth values that are in-painted. We mixed 2000 NYU image pairs from 16 scenes with 2000 pairs from Hypersim \cite{roberts2021hypersim}, which comes from 44 scenes and preserves sharp depth edges, and shuffled them. The synthetic test datasets use the eight Middlebury stereo scenes \cite{Middlebury2002}. In both cases, the HR depth and intensity maps have GT, and the LR input is synthesized to match the VL53L9's $54\times42$ zone geometry. This synthesis step is important to emulate the real sensor's image modality. A conventional imresize with bilinear or bicubic interpolation anti-aliases when downsampling HR to LR images. The kernel window spans several pixels and blurs zone boundaries from HR images. In reality, each SPAD zone reports one absolute depth value without averaging. Each zone generates its own timing histogram and reports a single distance. A network trained on over-smoothed LR input learns soft transitions and degrades reconstruction fidelity. Therefore, we replace the interpolation with an explicit zone model. For each zone, we reduce the corresponding block of GT pixels by winner-take-all (WTA) on the return strength $w = {I}/{d^{2}}$ \cite{martin2022_oe_video}, which is proportional to the number of photons from a surface with reflectance $I$ at distance $d$. The zone reports the depth of the pixel maximizing $w$, mimicking a receiver sensing two surfaces within the same illumination cone, where the stronger return determines the measured distance. Its behavior better matches dToF transmitters to detect peaks. Intensity is simply averaged over the HR pixels in one zone. The WTA mechanism can be described as
\begin{equation}
d_{\mathrm{LR}}(m,n) = d_{\mathrm{HR}}
\left(\arg\max_{(u,v)\in\Omega_{mn}}
\frac{I_{\mathrm{HR}}(u,v)}{
d_{\mathrm{HR}}^{2}(u,v)}
\right),
\label{eq:zone}
\end{equation}
where $\Omega_{mn}$ is the HR samples subtended by zone $(m,n)$. 
\section{Results}
\subsection{Evaluation on Synthetic Data}
\begin{figure*}[t]
  \centering
  \includegraphics[width=\textwidth]{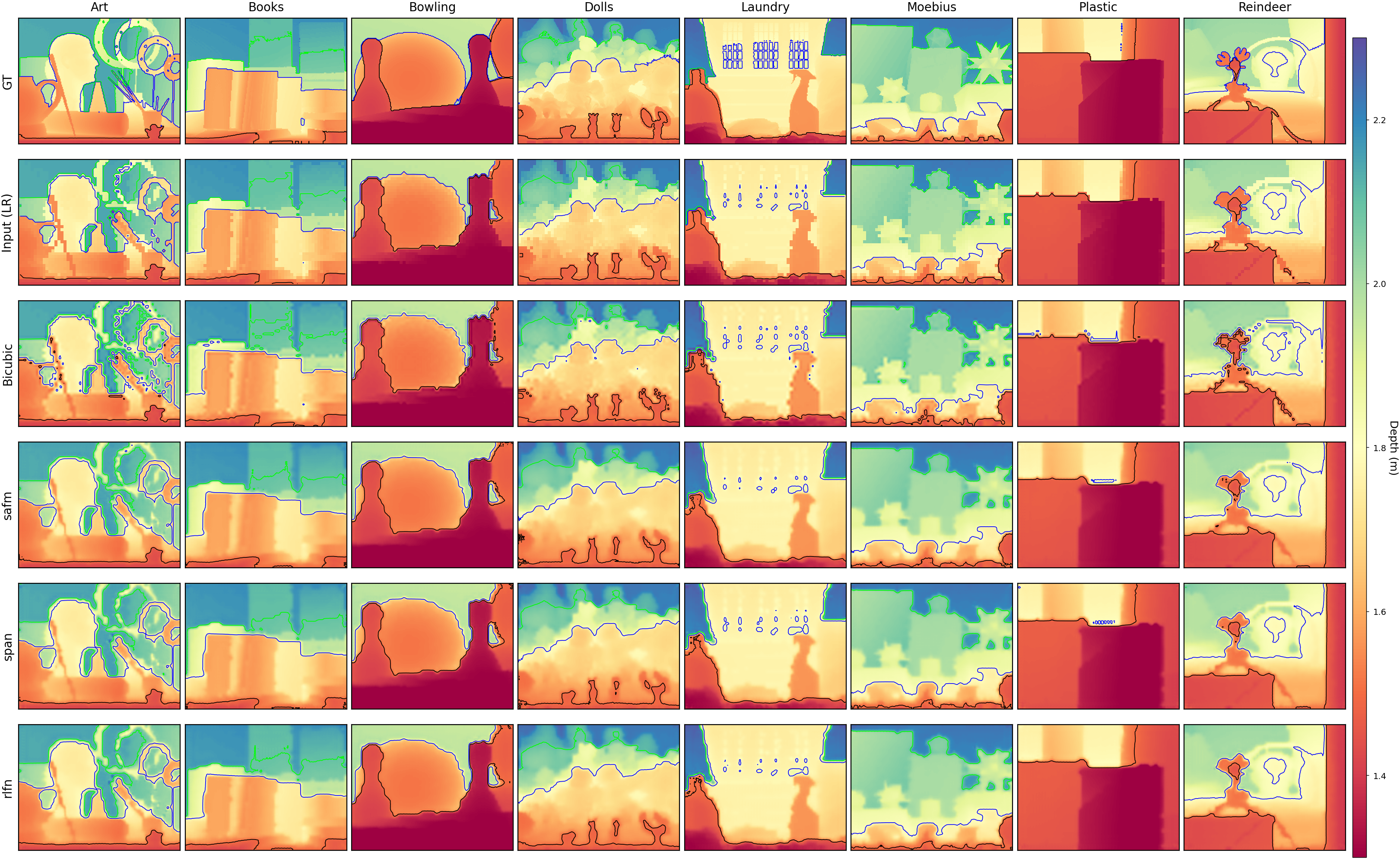}
  \caption{Depth SR on the Middlebury test scenes, $54\times42 \rightarrow 216\times168$ ($\times4$). Rows, top to bottom: ground truth (GT), VL53L9 LR input, bicubic upsampling, and the SAFM, SPAN, and RLFN backbones. Black, blue, and green curves are iso-depth contours at \SI{1.5}{\meter}, \SI{1.8}{\meter}, and \SI{2.1}{\meter}; their agreement with the GT row indicates whether surfaces are placed at the correct distance. The model size parameters are \#FC=12, \#B=1.}
  \label{fig:compare_depth}
\end{figure*}

\begin{figure*}[t]
  \centering
  \includegraphics[width=\textwidth]{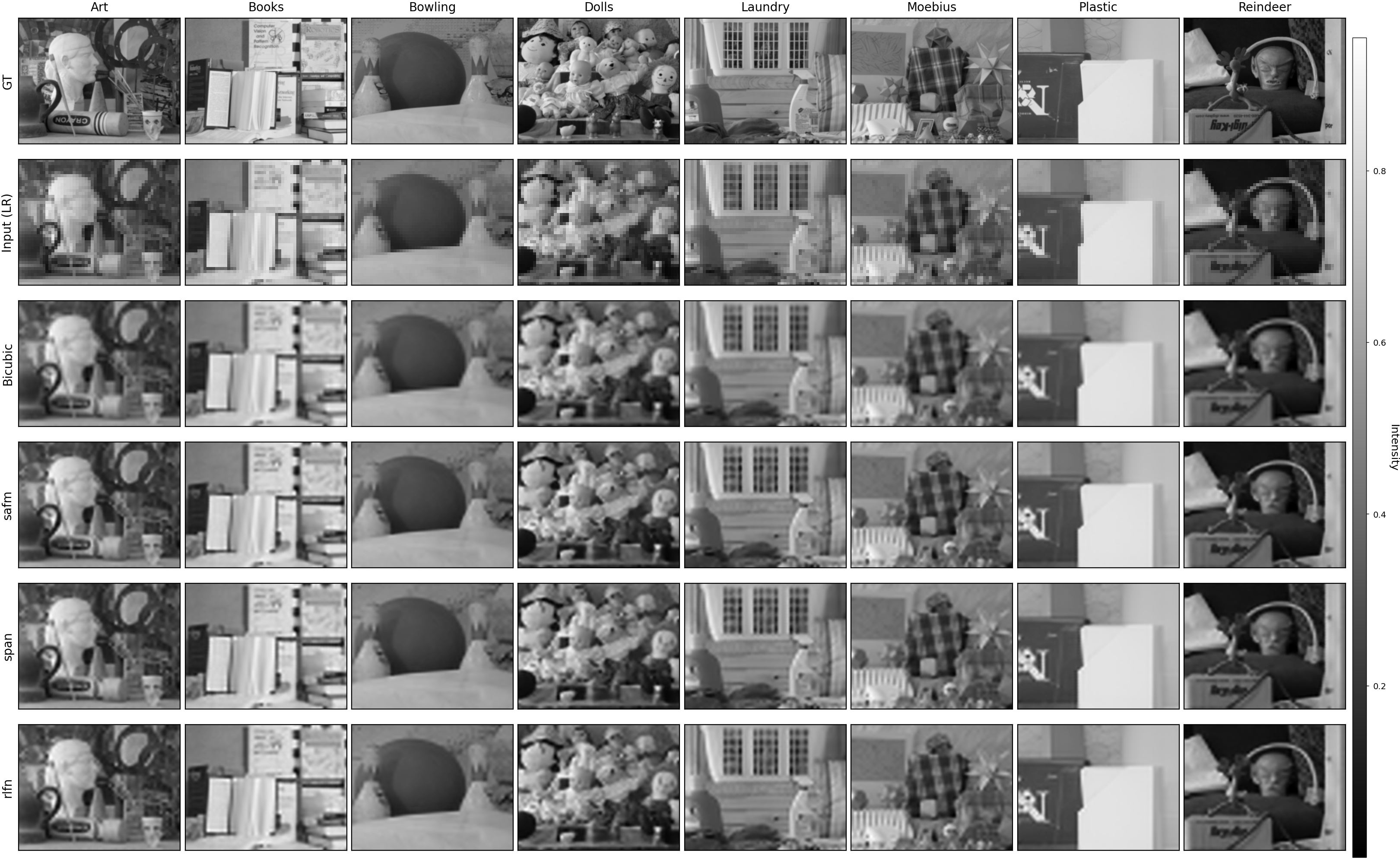}
  \caption{Intensity SR on the Middlebury test scenes, $54\times42 \rightarrow 216\times168$ ($\times4$). Rows, top to bottom: ground truth (GT), VL53L9 LR input, bicubic upsampling, and the SAFM, SPAN, and RLFN backbones. Intensity values are normalized. The model size parameters are \#FC=12, \#B=1.}
  \label{fig:compare_intensity}
\end{figure*}

\begin{table}[t]
  \centering
  \footnotesize
  \setlength{\tabcolsep}{3.0pt}
  \renewcommand{\arraystretch}{0.9}
  \caption{Depth and intensity reconstruction on the Middlebury test scenes,
           $54\times42 \rightarrow 216\times168$, averaged over eight scenes.}
  \label{tab:recon}
  \begin{tabular}{lcccc}
    \toprule
    \multicolumn{5}{c}{\emph{Depth}} \\
    \midrule
    Method & $\delta_1\uparrow$ & $\delta_2\uparrow$ & RMSE (m)$\downarrow$ & AbsRel$\downarrow$ \\
    \midrule
    Bicubic & 0.996 & 1.000 & 0.051 & 0.011 \\
    SAFM    & \textbf{0.998} & 1.000 & \textbf{0.042} & \textbf{0.009} \\
    SPAN    & \textbf{0.998} & 1.000 & \textbf{0.042} & 0.010 \\
    RLFN    & \textbf{0.998} & 1.000 & \textbf{0.042} & \textbf{0.009} \\
    \midrule
    \multicolumn{5}{c}{\emph{Intensity}} \\
    \midrule
    Method & PSNR$\uparrow$ & SSIM$\uparrow$ & GMSD$\downarrow$ & E-PSNR$\uparrow$ \\
    \midrule
    Bicubic & 26.88 & 0.7952 & 0.0017 & 10.49 \\
    SAFM    & \textbf{27.71} & \textbf{0.8226} & \textbf{0.0014} & \textbf{11.56} \\
    SPAN    & 27.56 & 0.8187 & \textbf{0.0014} & 11.40 \\
    RLFN    & 27.52 & 0.8161 & \textbf{0.0014} & 11.34 \\
    \bottomrule
    \multicolumn{5}{l}{\footnotesize $^{a}$ $\delta_3$ reaches $1.000$ for every method and is omitted.} \\
  \end{tabular}
\end{table}

\begin{table}[t]
  \centering
  \footnotesize
  \setlength{\tabcolsep}{3.0pt}
  \renewcommand{\arraystretch}{0.9}
  \caption{Reconstruction results of SAFM full and compressed models on the
           Middlebury test scenes, $54\times42 \rightarrow 216\times168$,
           averaged over eight scenes.}
  \label{tab:precision}
  \begin{tabular}{lcccc}
    \toprule
    \multicolumn{5}{c}{\emph{Depth}} \\
    \midrule
    Method & $\delta_1\uparrow$ & $\delta_2\uparrow$ & RMSE (m)$\downarrow$ & AbsRel$\downarrow$ \\
    \midrule
    Bicubic     & 0.996 & 1.000 & 0.0514 & 0.0110 \\
    Torch FP32  & \textbf{0.998} & 1.000 & \textbf{0.0423} & \textbf{0.0093} \\
    ONNX FP32   & \textbf{0.998} & 1.000 & \textbf{0.0423} & \textbf{0.0093} \\
    ONNX INT8   & \textbf{0.998} & 1.000 & 0.0470 & 0.0158 \\
    \midrule
    \multicolumn{5}{c}{\emph{Intensity}} \\
    \midrule
    Method & PSNR$\uparrow$ & SSIM$\uparrow$ & GMSD$\downarrow$ & E-PSNR$\uparrow$ \\
    \midrule
    Bicubic     & 26.88 & 0.7952 & 0.0017 & 10.49 \\
    Torch FP32  & \textbf{27.71} & \textbf{0.8226} & \textbf{0.0014} & \textbf{11.56} \\
    ONNX FP32   & \textbf{27.71} & \textbf{0.8226} & \textbf{0.0014} & \textbf{11.56} \\
    ONNX INT8   & 27.68 & 0.8213 & \textbf{0.0014} & 11.54 \\
    \bottomrule
  \end{tabular}
\end{table}

Table~\ref{tab:recon} reports the reconstruction performance of the three backbones and bicubic interpolation, averaged over the eight Middlebury scenes. All three learned models reduce depth RMSE from \SI{0.051}{\ meter} for bicubic interpolation to \SI{0.042}{\ meter}, an improvement of $\sim$18\%. The $\delta$ thresholds are nearly saturated for all methods, indicating that these metrics provide limited discrimination over the \SIrange{0.4}{2.3}{\metre} depth range considered here. The separation is clearer for intensity reconstruction. SAFM achieves the highest PSNR, SSIM, and E-PSNR, improving PSNR by \SI{0.83}{\decibel} and E-PSNR by \SI{1.07}{\decibel} over bicubic interpolation. All three learned models obtain the same rounded GMSD of 0.0014, compared with 0.0017 for bicubic interpolation. We also report gradient magnitude similarity deviation (GMSD)~\cite{xue2013gradient}, the standard deviation of gradient magnitude similarity between the reconstruction and the reference, which is sensitive to edge structure rather than absolute intensity (lower is better). SAFM obtains the lowest value. Fig.~\ref{fig:compare_depth} shows the HR depth reconstruction. All three networks remove the zone quantization visible in the LR input row, as does bicubic; the networks differ in boundary placement rather than overall error. Black, blue, and green curves mark where the reconstructed depth crosses 1.5, 1.8, and 2.1,m. Each curve is a level set of the depth map. Where a curve departs from its counterpart in the GT row, that surface has been placed at the wrong distance, which is an error the color map alone can hide. In \emph{Art}, the thin rods and the ring outlines are observable as continuous structures, whereas bicubic breaks them into the isolated blobs of the zones. In \emph{Reindeer}, parts of the antlers become spatially merged into a homogeneous region in the LR input, making their separation unobservable from the measurement. Consequently, the SR network cannot reliably recover the original split geometry. The intensity results in Fig.~\ref{fig:compare_intensity} are more visually distinct. The text in \emph{the} books, the window mullions in \emph{the} laundry, and the plaid pattern in \emph{Moebius} are recovered with better legibility than Bicubic, which agrees with the metric gaps in that half of Table~\ref{tab:recon}. Using the most accurate model, SAFM, we quantify quantization errors in Table~\ref{tab:precision} by comparing the full-precision SAFM model with the post-quantized INT8 ONNX model converted to the C code. The PyTorch FP32 and ONNX FP32 models produce identical results, confirming that ONNX export has no reconstruction error. INT8 quantization has little effect on intensity reconstruction, where PSNR decreases from 27.71 to 27.68~dB and SSIM from 0.8226 to 0.8213. The effect also occurs for depth, where RMSE increases slightly from \SI{0.0423}{\metre} to \SI{0.0470}{\metre} and AbsRel increases from 0.0093 to 0.0158, and $\delta_1$ and $\delta_2$ remain unchanged.

\subsection{Evaluation on Real Data}
\begin{figure*}[t]
  \centering
  \includegraphics[width=\textwidth]{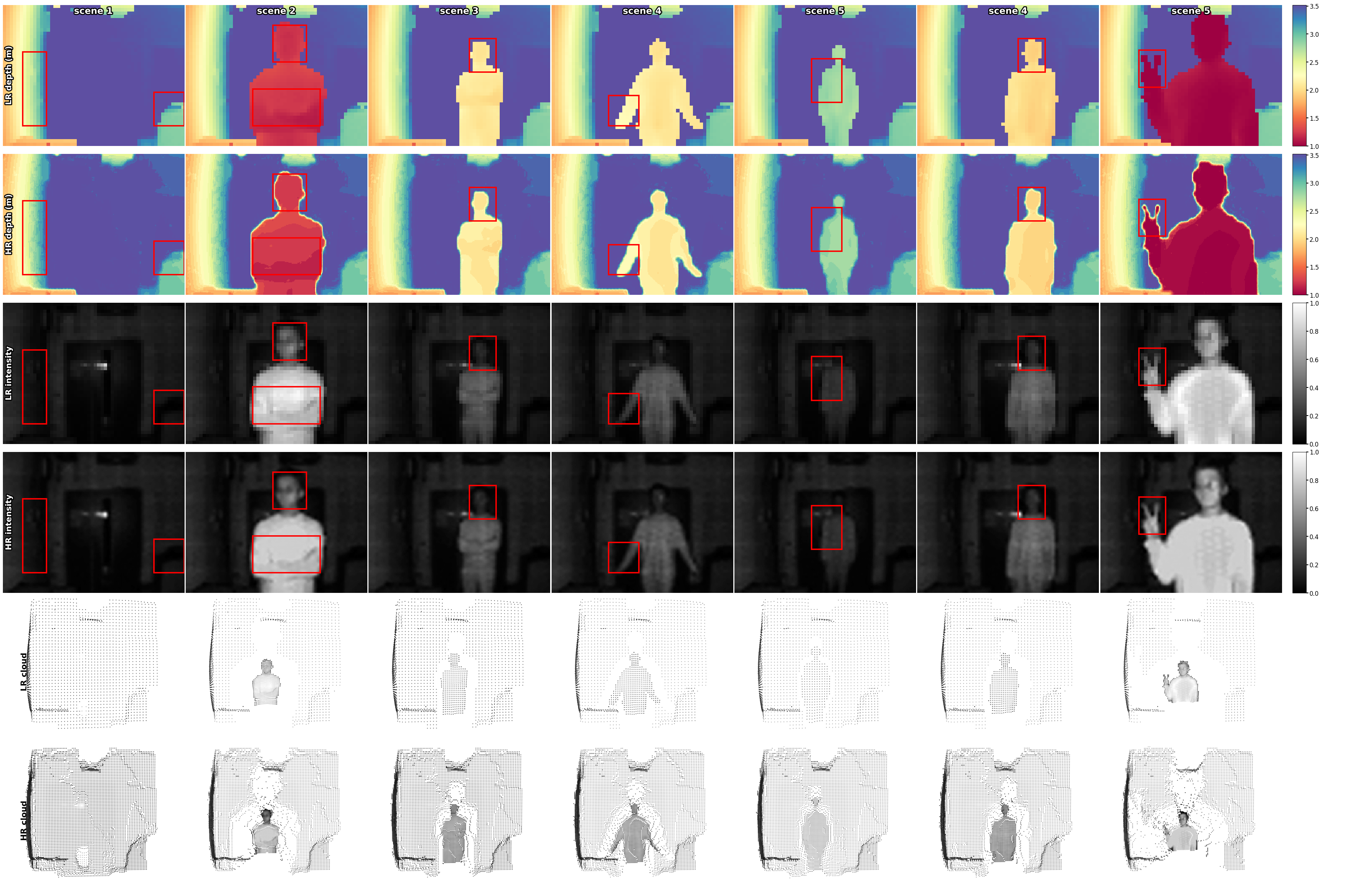}
    \caption{On-device $\times4$ SR of VL53L9 captures. Rows: LR depth ($54\times42$), HR depth ($216\times168$), LR and HR intensity, and the LR and HR point clouds. Red boxes mark the same region in the LR and HR panels of each column to directly compare LR and HR pairs. Depth in meters on a shared color scale; intensity normalized to $[0,1]$. The model size parameters are \#FC=12, \#B=1.}
  \label{fig:real_scenes_short}
\end{figure*}
Figure~\ref{fig:real_scenes_short} shows results in seven captures: an empty scene and a person walking from close to far and back. The first column contains no subject and shows only a wall and a door. The LR depth gradient across the side wall is not smooth but broken into steps, whereas at HR the same gradient is resolved without the striping. In the remaining columns, the subject is a coarse silhouette with jagged edges in LR, while in the HR images the outline, the gap between arm and torso, and the raised hand are clearly resolved. The intensity channel improves similarly: facial features and clothing texture emerge from what LR renders as flat bright patches, and the boundary between the subject and the dark background sharpens, although the photometric levels remain unchanged. The point clouds show the same 3D depth information, colored by intensity: the LR cloud shows the subject in a few depth steps with sparse points, whereas the HR cloud separates the person from the wall behind.
\section{Discussion}
\subsection{Ablation Study}
\begin{figure}[t]
  \centering
  \includegraphics[width=\columnwidth]{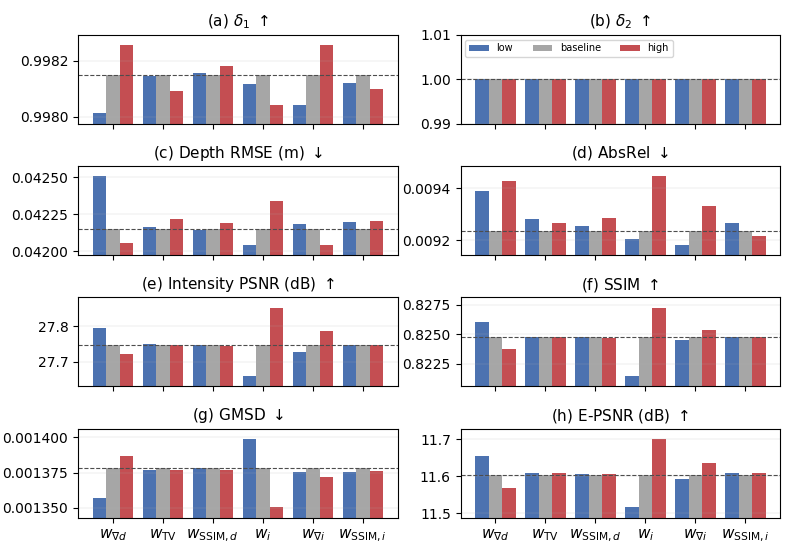}
  \caption{Loss-weight ablation on SAFM, averaged over the eight Middlebury test scenes. Each group varies one weight while holding the others at their default values; the grey bar and dashed line mark the baseline run, which uses the same model across groups. Settings are $w_{\nabla d}$ $(0.05, 0.2, 0.5)$, $w_{\mathrm{TV}}$ $(0, 10^{-4}, 10^{-3})$, $w_{\mathrm{SSIM},d}$ $(0.01, 0.05, 0.2)$, $w_{i}$ $(0.3, 1, 3)$, $w_{\nabla i}$ $(0.01, 0.05, 0.2)$, $w_{\mathrm{SSIM},i}$ $(0.03, 0.1, 0.3)$. The model size parameters are \#FC=12, \#B=1.}
  \label{fig:ablation}
\end{figure}
We conducted an ablation study of loss weighting to assess how much the chosen weights affect reconstruction quality. We combine six weighted terms, so we train the SAFM model twelve times, each run varying one weight by roughly a factor of four in each direction while holding the other five at the baseline values. All runs share the training settings and are evaluated on the same eight Middlebury scenes. Fig.~\ref{fig:ablation} shows that varying $w_{\mathrm{TV}}$, $w_{\mathrm{SSIM},d}$ and $w_{\mathrm{SSIM},i}$ over an order of magnitude does not change the performance. The other three affect the performance marginally. $w_{\nabla d}$ and $w_i$ show the largest variations. Increasing $w_{\nabla d}$ from 0.2 to 0.5 increases both depth RMSE and AbsRel, indicating a trade-off between edge emphasis and absolute depth accuracy. Increasing $w_i$ to 3.0 improves several metrics in this one-factor-at-a-time sweep, while $w_{\nabla d}=0.5$ gives the lowest intensity PSNR, SSIM, and E-PSNR. GMSD behaves differently because it directly rewards gradient similarity and therefore improves for sharper reconstructions even when pixel-wise metrics degrade. Overall, the changes remain small, with intensity PSNR spanning approximately \SI{0.15}{\decibel}. The sweep suggests that $w_{\nabla d}=0.05$, $w_i=3.0$, and $w_{\nabla i}=0.2$ are promising candidates for further joint tuning.

\begin{table}[t]
  \centering
  \footnotesize
  \setlength{\tabcolsep}{3.0pt}
  \renewcommand{\arraystretch}{0.9}
  \caption{Depth and intensity reconstruction averaged results on the Middlebury datasets, $54\times42 \rightarrow 216\times168$; the first two convolutional modules between the two branches are shared.}
  \label{tab:recon_share}
  \begin{tabular}{lcccc}
    \toprule
    \multicolumn{5}{c}{\emph{Depth}} \\
    \midrule
    Method & $\delta_1\uparrow$ & $\delta_2\uparrow$ & RMSE (m)$\downarrow$ & AbsRel$\downarrow$ \\
    \midrule
    Bicubic & 0.996 & 1.000 & 0.051 & \textbf{0.011} \\
    SAFM    & \textbf{0.998} & 1.000 & \textbf{0.045} & 0.012 \\
    SPAN    & \textbf{0.998} & 1.000 & 0.048 & 0.012 \\
    RLFN    & \textbf{0.998} & 1.000 & \textbf{0.045} & \textbf{0.011} \\
    \midrule
    \multicolumn{5}{c}{\emph{Intensity}} \\
    \midrule
    Method & PSNR$\uparrow$ & SSIM$\uparrow$ & GMSD$\downarrow$ & E-PSNR$\uparrow$ \\
    \midrule
    Bicubic & 26.88 & 0.7952 & 0.0017 & 10.49 \\
    SAFM    & \textbf{27.63} & \textbf{0.8206} & \textbf{0.0014} & \textbf{11.47} \\
    SPAN    & 27.28 & 0.8090 & 0.0015 & 11.13 \\
    RLFN    & 27.46 & 0.8161 & 0.0015 & 11.31 \\
    \bottomrule
    \multicolumn{5}{l}{\footnotesize $^{a}$ $\delta_3$ reaches $1.000$ for every method and is omitted.} \\
  \end{tabular}
\end{table}

A second ablation shares the two convolutional modules before the backbone, so these modules run only once per frame. As shown in Table~\ref{tab:recon_share}, depth reconstruction degrades for all three backbones: RMSE increases from \SI{0.042}{\metre} to \SI{0.045}{\metre} for SAFM and RLFN, and to \SI{0.048}{\metre} for SPAN. AbsRel increases from 0.009 to 0.012 for SAFM, from 0.009 to 0.011 for RLFN, and from 0.010 to 0.012 for SPAN. Intensity PSNR decreases by 0.08~dB for SAFM, 0.06~dB for RLFN, and 0.28~dB for SPAN. Host-GPU inference latency is reduced to \SI{0.48}{\milli\second}, \SI{0.31}{\milli\second}, and \SI{0.41}{\milli\second} for SAFM, SPAN, and RLFN, respectively. Sharing the stem therefore improves execution speed at the cost of reconstruction accuracy, with SPAN exhibiting the largest accuracy degradation.

\subsection{Design Constraint and Potential Improvement}
This work does not aim to introduce a new model architecture. Instead, we adopt a modular framework that accommodates emerging lightweight backbones, simplifies the conventional PixelShuffle-based reconstruction head, and optimizes the deployment pipeline at the system level. Further improvements could therefore be achieved not only by developing more efficient backbone architectures, but also by optimizing the compilation and execution pipeline. As ST Edge AI currently compiles the model into an optimized C implementation through a predefined deployment interface, lower-level optimizations, such as improvements to specific neural network kernels, more effective exploitation of target-specific DSP instruction sets, and optimized activation and intermediate-buffer management, could further reduce both inference latency and peak memory consumption. In addition, depth images can contain large background regions where HR reconstruction provides little useful information. Applying SR uniformly to such regions incurs unnecessary computation and increases latency. A region-of-interest (ROI)-based SR strategy could reduce this overhead by restricting HR reconstruction to spatial regions that require greater detail. Despite the computational savings, the ROIs could be based on predefined criteria, sensor-derived information, or a learned ROI detector. Data-driven DL requires sufficiently annotated training data. The effectiveness of ROI-based SR should therefore be evaluated on both ROI-selection accuracy and computational savings.

\section{Conclusion}
This work investigates on-device SR of LR depth and intensity images from a compact VL53L9CX SPAD imager under strict memory and computational constraints. The proposed framework combines lightweight SR backbones with a memory-efficient two-stage PixelShuffle reconstruction head and a simplified sensor-processing pipeline, enabling deployment on an STM32H563ZI MCU with only \SI{640}{\kibi\byte} of SRAM. We evaluate three lightweight backbones on reconstruction accuracy, computational cost, and deployment memory. INT8 quantization preserves intensity reconstruction closely but introduces a measurable degradation in absolute depth accuracy. Synthetic experiments demonstrate consistent improvement over bicubic reconstruction, and real measurements show substantially denser depth, intensity, and point-cloud outputs. The resulting system performs SR entirely on a bare-metal MCU without requiring a GPU, embedded Linux processor, or dedicated neural processing unit.


\bibliographystyle{IEEEtran}
\bibliography{references}

\end{document}